\documentclass[12pt,letterpaper,onecolumn]{quantumarticle}
\pdfoutput=1
\usepackage{amsmath,amsthm,amssymb,tikz,tikz-cd,braket,comment,float,graphics}
\usepackage[normalem]{ulem}
\usetikzlibrary{arrows,trees,decorations.markings,patterns}
\usepackage{wrapfig}
\newtheorem{proposition}{Proposition}
\usepackage[
colorlinks=true,
linkcolor=blue,
citecolor=red,
urlcolor=blue
]{hyperref}
\usepackage[
backend=biber,
style=numeric-comp,
sorting=none,
maxbibnames=2,
minbibnames=1,
maxcitenames=2,
mincitenames=1,
giveninits=true,
terseinits=true
]{biblatex}
\AtEveryBibitem{
\clearfield{issn}
\clearfield{doi}
\clearfield{url}
}
\renewbibmacro*{volume+number+eid}{%
\printfield{volume}%
\printfield[parens]{number}%
\printfield{eid}}
\usepackage[normalem]{ulem}

\usepackage{xcolor}

\usepackage{xcolor}

\usepackage{xcolor}

\begin{document}

\title[Quantum Advantage over Wirings of Nonsignaling Boxes in Multipartite Networks]{Quantum Advantage over Wirings of Nonsignaling Boxes in Multipartite Networks}
\author{Peter Bierhorst}
\author{Arkaprabha Ghosal}
\author{Soumyadip Patra}




\begin{abstract}
Quantum-entangled measurements are known to enable multi-party behaviors that are impossible with unentangled measurements on nonlocal resources, even those that are super-quantum and bound only by the no-signaling principle. This advantage can be witnessed by the entanglement swapping protocol, along with corresponding impossibility results for ``nonlocality swapping''. However, the advantage assumes the absence of pre-existing nonlocal resources shared by the swapped-to parties; it no longer holds if all pairs of parties are allowed to share bipartite nonlocal resources. Here, we consider a resource-theoretic perspective in which bipartite nonclassical resources are free resources that can be shared by any pair of parties in a multipartite network, and ask whether quantum entangled measurements can still provide an advantage over certain basic measurements, known as wirings, of nonsignaling nonlocal resources. 
We resolve this question in the affirmative by demonstrating an explicit four-party behavior that \textit{can} be achieved with bipartite quantum resources subject to entangled measurements, and \textit{cannot} be achieved if the bipartite resources are allowed to be more general nonsignaling nonlocal ``boxes'' so long as the measurements are restricted to local wirings, even also allowing for globally shared classical randomness. Furthermore, the argument generalizes: the same separation can be witnessed for $K+2$ parties with access to $K$-partite nonlocal resources for any $K > 2$. We also examine these results in different contexts, such as the star network configurations and scenarios not admitting globally shared classical randomness, further enhancing understanding of the capabilities of entangled measurements in multi-party configurations.
\end{abstract}


\section{Introduction}

Quantum-entangled measurements will form a central component of quantum networks, which hold the promise to transform communication by enabling a quantum-secured internet. The importance of entangled measurements to quantum networks is exemplified by the quantum repeater~\cite{Azuma2023}, a fundamental network component powered by entangled measurements via the protocol of entanglement swapping \cite{bennett:1993}. Furthermore, multi-party entangled measurements, such as measurements in the Greenberger-Horne-Zeilinger (GHZ) basis~\cite{GHZ}, facilitate the creation and distribution of multi-party entangled states for use in protocols such as efficient distribution of cryptographic key to multiple parties~\cite{epping:2016}.

\begin{wrapfigure}{r}{0.45\textwidth}
\vspace{-8mm}
\includegraphics[scale=0.95]{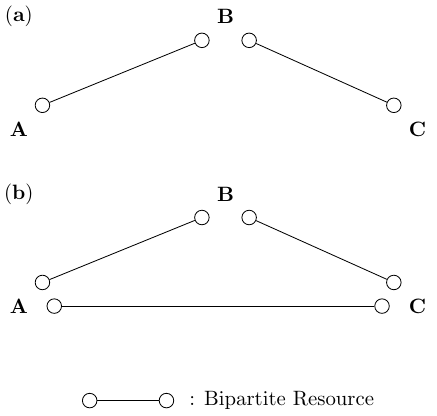}
\caption{Three-party networks of parties Alice (A), Bob (B), and Charlie (C) where underlying bipartite nonclassical resources are available in (a) a star network with B at the center, and (b) a fully connected configuration in which all pairs have access to a bipartite nonclassical resource.}\label{f:tripart}
\end{wrapfigure}

In quantum theory, the enhanced capabilities of entangled measurements over non-entangled measurements can be characterized in a direct and fundamental manner: if three parties Alice, Bob, and Charlie are arranged as in Figure \ref{f:tripart} (a) such that Alice and Bob share a bipartite entangled state, and Bob and Charlie share an independent bipartite entangled state, but no other pre-existing entanglement exists, then a non-entangled measurement at Bob \textit{cannot} result in Alice and Charlie sharing entanglement, whereas an entangled measurement can entangle Alice and Charlie's system conditioned on Bob's measurement outcome via entanglement swapping. This advantage can be formulated in a more general, theory-independent manner by considering bipartite resources bound only by the no-signaling condition, i.e.~potentially super-quantum  resources like the Popescu-Rohrlich (PR) ``box"~\cite{PRBOX}: even if Alice-Bob and Charlie-Bob are allowed any number of such resources, it is impossible (see, for instance,~\cite{short:2006} Sec.~IV,~\cite{chao:2017} Thm.~4.2.2, and~\cite{Short_2010} Thm.~5 to replicate the entanglement-swapping behavior when Bob's measurement of his subsystems comprises only classical manipulations, i.e., ``wirings'' that treat each measured subsystem strictly as a black box into which an input can be supplied and from which an output recorded, while allowing for recorded outputs to be used as inputs to other measured boxes. These impossibility results demonstrate the strength of entangled measurements in a theory-agnostic framework: quantum-entangled measurements can confer an advantage to any no-signaling theory, even though such theories permit, in principle, stronger-than-quantum correlations.

Intriguingly, the existence of such a theory-agnostic advantage has not been demonstrated if we consider a fully connected network in which each party pair has access to a bipartite resource---that is, the scenario of the previous paragraph, but where additionally Alice and Charlie now have access to a bipartite entangled state; see Figure \ref{f:tripart} (b). Consideration of such fully connected networks is motivated by a resource-theoretic perspective on the question  of whether bipartite quantum resources (with entangled measurements) confer an advantage over basic measurements of general nonsignaling bipartite resources: if we consider bipartite resources to be the ``free" resources of the scenario, they should be available to any pair of parties. Here it is reasonable to also allow the more basic resource of globally shared classical randomness. In the fully symmetric case of Figure~\ref{f:tripart} (b), the entanglement swapping scenario of the previous paragraph now turns out to be easy to spoof without entangled measurements by exploiting shared classical randomness, as described in the context of quantum resources in the comments preceding Eq.~(16) in~\cite{bancal:2018} and for more general nonsignaling resources in~\cite{chao:2017} (see Thm.~4.2.3). 

Now, there is a different behavior in the Figure \ref{f:tripart} (b) scenario, realizable if one party makes an entangled measurement, that is provably impossible to replicate if all parties are restricted to product measurements on underlying bipartite \textit{quantum} resources, even allowing bipartite quantum resources to be shared among all pairs of parties and allowing classical randomness to be globally shared among the three parties~\cite{bierhorst:2023}. In parallel with the entanglement-swapping scenario, it might be expected that allowing for more general nonsignaling resources would not help in replicating this behavior. However, it was shown in \cite{bierhorst:2023} that a network of (bipartite) PR-boxes measured with only local wirings can reproduce this behavior, and that such networks can even replicate a more intricate behavior of Ref.~\cite{rabelo:2011} that was constructed to witness an entangling measurement---that is, one that results in two remote parties becoming newly entangled conditioned on the measurement's outcome---\textit{fully} device-independently (i.e., even without assuming the absence of a three-way entangled resource, and only presuming an underlying quantum mechanical description).

This leaves open the question of whether entangled measurements on bipartite resources can outperform (locally) wired bipartite nonsignaling boxes in any multi-party scenario, if no restrictions are placed on which pairs of parties can possess the bipartite boxes. The question has foundational implications for how we understand entangled measurements: whether they unlock new fundamental capabilities of bipartite nonclassical resources beyond what is contraindicated only by these resources' no-signaling character, or if these new capabilities are a specifically quantum feature, dependent on quantum limitations of the entangled states such as Tsirelson's bound~\cite{tsirelson:1993}. 

In this paper, we answer the question by demonstrating the following: there is a behavior in a four-party network that can be witnessed by entangled measurements on an underlying network of bipartite-only quantum resources, that \textit{cannot} be witnessed with measurements on an underlying network of bipartite nonsignaling resources (possibly super-quantum), provided the measurements are restricted to local wirings, even in the presence of globally shared classical randomness. To obtain this behavior, one party applies a GHZ-basis measurement to their portion of three Bell states shared independently with the other three parties to swap a tripartite GHZ state to the other parties; then conditioned on success of the swap, those parties can witness genuine tripartite nonlocality according to the local-operations-shared-randomness (LOSR) paradigm \cite{coiteux:2021}. We prove that it is impossible to replicate the resulting behavior with local wirings and bipartite nonsignaling resources, using the formal techniques of Ref.~\cite{bierhorst:2024}. Our demonstration is noise-robust and so admits the possibility of an experimental test, which would be device-independent in character in the sense that the conclusions can be drawn directly from the experimentally observed statistics without detailed assumptions about the inner workings of measuring devices and measured states.

We also show the argument can be generalized to prove an analogous separation for networks of $K+2$ parties sharing at most $K$-partite nonsignaling resources for any integer $K\ge 2$. Notably, in the generalization our quantum-achievable behavior only ever requires bipartite resources (as opposed to the ostensibly available $K$-partite resources), and only uses those arranged in a star network. We thus explore the implications of our results in more restrictive paradigms that bar more-than-bipartite resources in star networks, while also examining paradigms barring globally shared classical randomness. Such other paradigms allow a more natural extension to $K=1$ and $K=0$ cases, and they provide a fuller perspective on the power of the quantum-achievable correlations.


Returning to the $K\ge 2$ cases, these all involve performing a multi-partite entangled GHZ-basis measurement on Bell states, suggesting that it may be specifically the \textit{multi-partite} character of the entangled measurement that enables the advantage over unentangled measurements. This suggests an intriguing direction for future work classifying the power of different types of entangled measurements in a theory-independent setting. Another interesting open question is whether the separation studied in this work exists in the case of $K+1$ parties sharing $K$-partite resources. This question has relevance to a conjecture in Sec.~V of Ref.~\cite{coiteux:2021a}, which posits (for the tripartite scenario) that allowing for generalized analogues of entangled measurements on bipartite nonsignaling resources can lead to behaviors inaccessible to local-wiring measurements on the same class of resources. Since quantum mechanics is an example of a theory with an ``analogue'' of entangled measurements (i.e., entangled measurements themselves), a positive answer to our question in a tripartite network ($K=2$) witnesses an affirmative answer to this conjecture---though the reverse implication would not necessarily hold. 

Finally, we remark that our results constraining behaviors realizable with basic measurements on potentially-exotic super-quantum nonsignaling resources, will also immediately constrain behaviors realizable with basic measurements on strictly \textit{quantum} resources. The use of entangled measurements to rule out only-local-wirings on quantum resources has received recent attention in contexts both allowing globally shared classical randomness \cite{bierhorst:2023} and disallowing it \cite{supic:2022}; constraints derived in the manner of this paper thus contribute to our fundamental understanding of entangled measurements and their capabilities across a broad class of more physical paradigms.

\section{Two Paradigms for Networked Nonsignaling Resources}\label{s:qem}
\begin{wrapfigure}{r}{0.4\textwidth}
\vspace{-0.3cm}
\begin{tikzpicture}[scale=1.1,every node/.style={circle, draw, minimum size=7pt, inner sep=2pt}, every path/.style={thick, draw=black}]
\node (A1) at (0,0) {};
\node (A2) at (-0.1,-0.4) {};
\node (A3) at (-0.5,-0.5) {};
\node[draw=none] at (-0.4,-0.1) {\textbf{A}};
\node (C1) at (-0.5,-4) {};
\node (C2) at (-0.1,-4.1) {};
\node (C3) at (0,-4.5) {};
\node[draw=none] at (-0.45,-4.353) {\textbf{C}};
\node (B1) at (4,0) {};
\node (B2) at (4.1,-0.4) {};
\node (B3) at (4.5,-0.5) {};
\node[draw=none] at (4.4,-0.1) {\textbf{B}};
\node (D1) at (4.5,-4) {};
\node (D2) at (4.1,-4.1) {};
\node (D3) at (4,-4.5) {};
\node[draw=none] at (4.4,-4.353) {\textbf{D}};
\draw (A1) -- (B1);
\draw (A3) -- (C1);
\draw (C3) -- (D3);
\draw (B3) -- (D1);
\draw (A2) -- (D2);
\draw (C2) -- (B2);
\node (cap1) at (0,-5.5) {};
\node (cap2) at (1,-5.5) {};
\node[draw=none] at (2.8,-5.5) {: Bipartite Resource};
\draw (cap1) -- (cap2);
\end{tikzpicture}
\vspace{-1.5cm}
\caption{A four-party network in which each of the (six) pairs of parties shares a bipartite nonclassical resource, but there are no higher-order multipartite entangled resources. Globally shared classical randomness, which may be present, is not depicted.}\label{f:fourparty}
\end{wrapfigure}
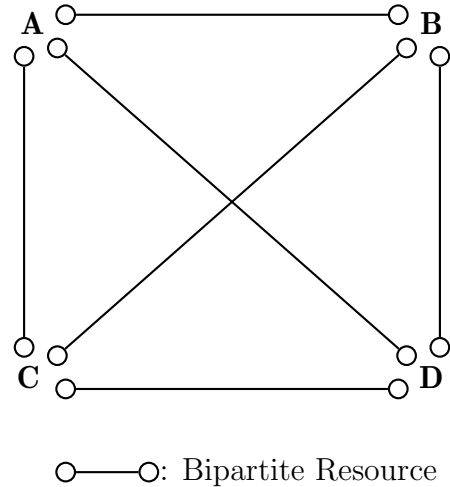

We start by considering multi-party networks in which we allow each pair of parties to share bipartite nonclassical resources, but no higher order nonclassical resources are available (e.g., three-party, four-party), though we may choose to allow for the possibility of globally shared classical randomness. This scenario is illustrated for four parties in Figure \ref{f:fourparty}.

We consider two paradigms: (1) \textsf{Q}uantum states, \textsf{E}ntangled \textsf{M}easurements (\textsf{QEM}): The scenario in which the nonclassical resources are quantum states, potentially entangled, and the parties are allowed to make entangled measurements on the resource portions they possess. (2) \textsf{N}o-\textsf{S}ignaling nonclassical ``boxes", local \textsf{W}irings (\textsf{NSW}): The scenario in which the nonclassical resources are ``boxes" with which the parties can interact in a black-box manner, i.e., by supplying inputs and recording outputs. These resources are bound only by the no-signaling condition and so can exhibit super-quantum bipartite correlations, like the Popescu-Rohrlich box~\cite{PRBOX}. They, however, do not admit ``entangled measurements" nor any generalized analogue thereof.

The \textsf{NSW} paradigm notably allows for the output of one box to be fed as input to another box possibly shared with a different player; this is the concept of ``local wirings" (see Section IIIC of \cite{barrett:2005} or also \cite{barrettpironio:2005,bierhorst:2023} for some examples). For now, we will consider globally shared classical randomness to be allowed as a ``free" resource in both paradigms. The main result is to describe a four-party behavior that can be induced
 in the \textsf{QEM} paradigm with bipartite resources, and prove that it cannot be produced in the \textsf{NSW} paradigm even allowing for globally shared classical randomness. 

Before describing this behavior, we provide some context by further reviewing some examples, briefly discussed in the introduction, that might seem to have been natural candidates to look for a \textsf{QEM}/\textsf{NSW} separation, but do not in fact witness it. First, the standard tripartite entanglement swapping experiment in the configuration of Figure \ref{f:tripart} (a) will not work, as it does not require an entangled measurement if we allow globally shared classical randomness and pre-shared entanglement between the swapped-to pair of parties (as shared between A and C in Figure \ref{f:tripart} (b)); see Ref.~\cite{bancal:2018,chao:2017}. This is because the swapped-to parties can measure the bipartite entangled resources they already possess, while using globally shared classical randomness to conspire with the swapping party to make it appear like their particular entangled resource depends on the swapping party's outcome, while in reality they just measure different pre-existing bipartite resources in different rounds with the choice based on the pre-existing shared classical randomness. Interestingly, such spoofing models do not even specifically require \textit{globally} shared classical randomness; bipartite classical randomness will suffice---we demonstrate this in Appendix \ref{s:spoofing}. So postselected entanglement or nonlocality between the swapped-to parties is not, by itself, a theory-agnostic witness of an entangled measurement at the swapping party. In the terminology of recent work~\cite{Ciudad_Alanon24}, such correlations remain in the ``shadow of Bell’s theorem'', insofar as their nonclassicality can be traced back to a postselected Bell violation between the swapped-to parties rather than to a genuinely new network effect. The next natural place to look for correlations witnessing a \textsf{QEM}/\textsf{NSW} separation would be tripartite quantum correlations that require entangled measurements even if all party pairs possess bipartite quantum resources, such as the behavior described in \cite{bierhorst:2023}, or the behavior of \cite{rabelo:2011} that device-independently witnesses an entangling measurement in a yet more general paradigm that does not even assume the absence of three-party quantum-entangled resources. Unfortunately, these behaviors can still be spoofed by \textsf{NSW} correlations \cite{bierhorst:2023}---in particular, spoofed with PR boxes and wirings; once again the spoofing models do not even require globally shared classical randomness. These examples illustrate how a $\textsf{QEM}\cap \textsf{NSW}^C$ witness is not so easy to find, even considering behaviors that otherwise require entangled measurements in very fundamental ways when only quantum resources are available.

\begin{figure}[!htb]
\includegraphics[scale=0.75]{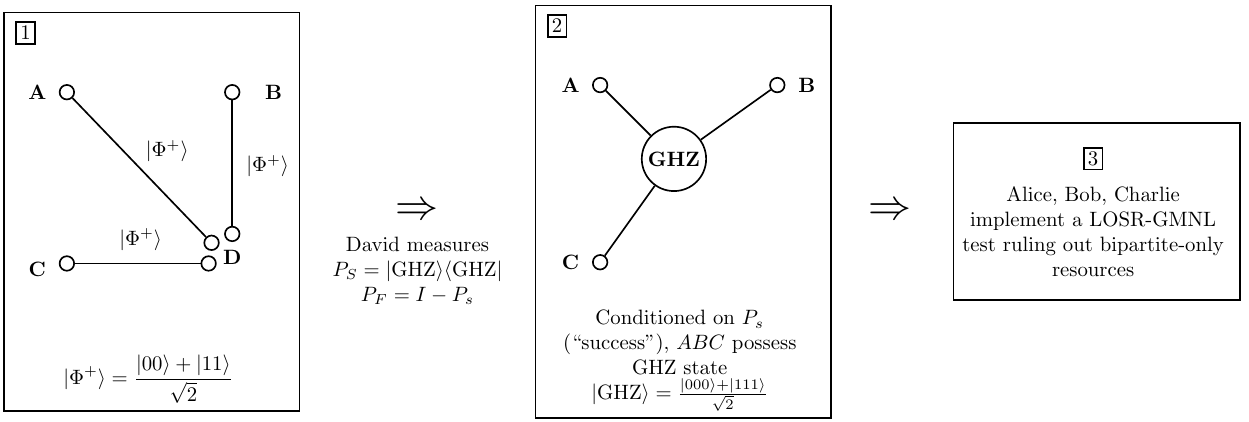}
\caption{Scheme for generating a behavior in the \textsf{QEM} paradigm, which will not be replicable in the \textsf{NSW} paradigm.}\label{f:scheme}
\end{figure}

To witness the separation, we employ the four-party scheme of Figure~\ref{f:scheme}. As indicated in the figure, each of the pairs $AD$, $BD$, $CD$ initially shares a copy of the Bell state $(\ket{00}+\ket{11})/\sqrt 2$; David thus possesses three qubits. While other bipartite nonclassical resources would be available in the \textsf{QEM} paradigm as depicted in Figure \ref{f:fourparty}, this scheme does not use them. To obtain the behavior, David performs on his three qubits a projection measurement onto the GHZ state $(\ket{000} + \ket{111})/\sqrt 2$; his outcome is either GHZ (success), or not-GHZ (failure). It is straightforward to compute that ``success" occurs with probability 1/8 and conditioned on success, Alice, Bob, and Charlie jointly possess a GHZ state (step 2 in the figure). Now, the GHZ state is a three-party resource, and Alice/Bob/Charlie can perform a test of ``Local Operations, Shared Randomness - Genuine Multipartite Nonlocality" (LOSR-GMNL) to rule out bipartite-only resources even in the presence of globally shared classical randomness, such as the test derived by Coiteux-Roy, Renou, and Wolfe \cite{coiteux:2021} or more recent improvements in Cao et al.~\cite{cao:2022} and Mao et al.~\cite{mao:2022}. For example, the Mao et al.~test of LOSR-GMNL is formulated for a tripartite scenario via an inequality that must be obeyed if only bipartite resources are present:
\begin{equation}\label{e:mao}
\langle A_0B_0\rangle +\langle A_0B_1\rangle+\langle A_1B_0C_1\rangle -\langle A_1B_1C_1\rangle +2\langle A_0C_0\rangle \le 4,
\end{equation}
which uses a standard notation where Alice, Bob, and Charlie each have settings $X$, $Y$, $Z$ taking values in the binary set $\{0,1\}$ and observe outcomes $A,B,C$ in the binary set $\{+1,-1\}$, and the expression $\langle A_0B_1\rangle$ (for example) is shorthand for the expected value
\begin{equation*}
\langle A_0B_1\rangle= \mathbb E(AB|X=0,Y=1)= \mathbb P(A=B|X=0,Y=1) -\mathbb P(A\ne B| X=0,Y=1).
\end{equation*}
While the overall marginal correlation $\mathbb P (A,B,C|X,Y,Z)$ cannot violate \eqref{e:mao} (because only bipartite resources are present), the conditional marginal correlation 
\begin{equation*}
\mathbb P_{\text{\footnotesize success}}(A,B,C|X,Y,Z) = \mathbb P(A,B,C|X,Y,Z; D=\textnormal{success})
\end{equation*}
will violate \eqref{e:mao} with the specific strategy given in Mao et al.~\cite{mao:2022} using the GHZ state.

Notably, the behavior is noise-robust in the sense that the states and measurements do not need to be perfect in order for the conditioned-on-success behavior to violate the inequality in~\eqref{e:mao}. The witnessing behavior is thus amenable to experimental test; as we shall see in the next section, \textsf{NSW} schemes are incapable of any success-conditioned violation of \eqref{e:mao}. We give a quantitative treatment of noise-robustness in Appendix \ref{s:resolution}, finding  that the inequality is violated if the fidelity of the measured Bell states exceeds $0.941$.

An interesting feature to consider is the straightforwardness of David's measurement approach: he does not have multiple different measurement settings, but rather always performs the same measurement which can only either succeed or fail. Alice, Bob, and Charlie, who have multiple settings, will yield data in the ``failure'' rounds that is uninteresting noise, as they are performing the same measurements regardless of success or failure of David's measurement (they have no way of knowing which David-outcome occurs in a given trial). We remark however it is possible to disambiguate David's measurement into a full eight-outcome GHZ basis measurement, in which each of his eight different outcomes project the $A$/$B$/$C$ state into a different GHZ-type state; we demonstrate in Appendix \ref{s:resolution} that each of these eight outcomes leads to a violation of a symmetry of inequality \eqref{e:mao} even with Alice/Bob/Charlie making the same measurements tailored to the canonical GHZ state. This is relevant for making the most effective use of all the generated data in a potential experimental realization, but since it does not otherwise enhance the theoretical implications of witnessing the \textsf{QEM}/\textsf{NSW} separation, we focus primarily on the conceptually simpler success/failure measurement scenario for David. 

\section{Impossibility Result For Behaviors Using \textsf{NSW} Paradigm}\label{s:imposs}
\begin{figure}[htbp]
\centering
\begin{minipage}[b]{0.47\textwidth}
\centering
\includegraphics[width=\linewidth]{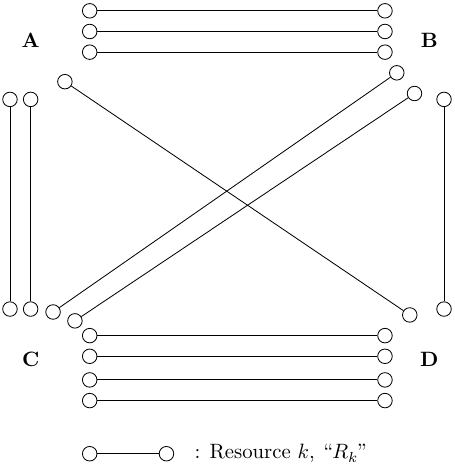}
\caption{The bipartite resources of Figure \ref{f:fourparty} can in general be multiple parallel resources measured in cascaded fashion, with outputs of resources fed as inputs to other resources.}\label{f:bundle}
\end{minipage}
\hfill
\begin{minipage}[b]{0.47\textwidth}
\centering
\raisebox{2.7ex}{%
\includegraphics[width=\linewidth]{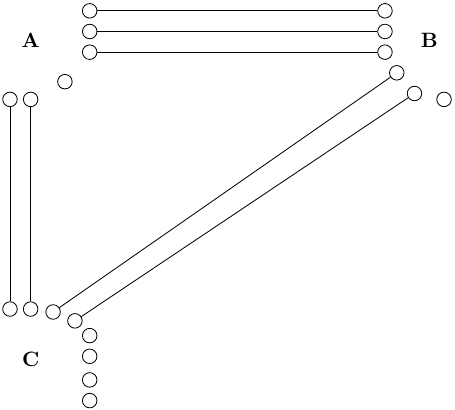}
}
\caption{
Conditioned on David measuring and receiving outputs from all of his bipartite resources, these resources that were previously shared with other parties as in Figure~\ref{f:bundle} reduce to unipartite local random variables.}\label{f:choppedbundles}
\end{minipage}
\end{figure}

We show it is impossible to replicate the behavior of Figure \ref{f:scheme} with a model obeying the \textsf{NSW} paradigm. To do so, let us start by describing the paradigm in a little more detail. First note that each bipartite \textsf{NSW} resource can be in general a bundle of parallel bipartite resources, as indicated in Figure~\ref{f:bundle}. In a measurement trial, each party receives a measurement setting; upon receiving this, they consult one of their bipartite resources, providing an input and getting an output, then move on to another bipartite resource potentially shared with a different player, and repeat. The choice of order in which later resources are measured, as well as the choice of inputs provided to them, can depend on outputs of earlier-measured resources, in addition to the party's measurement setting. When a player has measured all of their resources, they report a final outcome as a function of the observed resource outputs. We emphasize a choice of terminology here that helps distinguish observed variables from latent ones: each party receives a \textit{setting} and reports a final \textit{outcome}, which are observed quantities in an experiment; in contrast, the underlying unobserved nonsignaling resources receive \textit{inputs} and produce \textit{outputs}.

 Some loose intuition for why the behavior of Figure \ref{f:scheme} now \textit{cannot} be achieved is that after David measures his resources, his bipartite resources are effectively collapsed to one-party resources possessed by non-David parties, which are indistinguishable from (unshared) local randomness. To illustrate, the network of Figure~\ref{f:bundle} after David's measurement looks like Figure~\ref{f:choppedbundles}, where the orphan circles---which were once connected by lines to $D$---represent the portions of the bipartite resources that can still be measured by Alice (or Bob, or Charlie), but conditioned on David's measurement, these formerly-bipartite resources are effectively each reduced to a local random variable. An $A$/$B$/$C$ correlation compatible with the configuration in Figure \ref{f:choppedbundles} cannot violate an inequality witnessing 3-way nonlocal resources such as that of Mao et al.~in equation \eqref{e:mao}.

The above argument is informal, and its incomplete reasoning can indeed lead to flawed conclusions if taken too far---we will provide an example of this in the next section. However, a rigorous proof of the impossibility can be formulated using a framework for analyzing \textsf{NSW} correlations laid out in a recent paper authored by one of us~\cite{bierhorst:2024}. We provide the full proof in Appendix~\ref{s:app}, whose approach we sketch as follows: first, assume any globally shared classical randomness takes on a fixed value; the constraint that will then be derived---that is, satisfaction of \eqref{e:mao} even when conditioning on David's ``success" outcome---will apply for any such fixed value, and therefore will also apply to mixtures over all possible values of the shared classical randomness. This allows us to effectively ignore the shared classical randomness for the rest of the argument. Now, the framework of \cite{bierhorst:2024} formulates \textsf{NSW} by letting $R(\cdot|\cdot)$ denote the input-conditional distribution of a single resource, so for instance $R(AC|XZ)$ could represent a single PR box shared between Alice and Charlie; in a network of multiple resources these can be indexed as $R_k(\vec p_k|\vec s_k)$ where each $\vec p_k$ and $\vec s_k$ represents a pair of outputs and inputs, respectively, for the two parties sharing that resource. Then the probability of some particular list of outputs for all players' resources, 
\begin{equation*}
\text{Prob}(\text{The resources take some fixed value}|\text{A choice of all players' measurement settings}),    
\end{equation*}
is always expressible as a product of the resource distributions $\prod_{k=1}^m R_k(\vec p_k|\vec s_k)$ where it is possible for some of the resource inputs $s_k$ to depend not just on the party's setting, but also on outputs of other resources $p_k$ via the local wirings---see Eq.~(9) of \cite{bierhorst:2024}. Applying this to the current problem, one shows that once David has measured all of his resources and received outputs, then conditioned on David's resource outputs, the probability of the raw resource outputs of Alice, Bob, and Charlie is equal to
\begin{equation*}
\left(\prod_{k\in \mathcal D^C} R_k(\vec p_k|\vec s_k)\right)\left(\prod_{k\in \mathcal D} R_k( p_k|  s_k)\right)
\end{equation*}
where $\mathcal D$ is the set of resources $R_k$ that involved David (now reduced to singleton resources) and $\mathcal D^C$ is the set of bipartite resources shared among Alice/Bob/Charlie. The above expression is the probability distribution of a network containing no 3-party resources, and so the joint distribution of final outcomes---which are functions of the resource outputs---cannot violate \eqref{e:mao}. This is because \eqref{e:mao} was shown in \cite{mao:2022}---following arguments of the form pioneered by \cite{coiteux:2021}, which introduced the LOSR-GMNL notion of genuine multipartite nonlocality---to be a constraint obeyed by all tripartite networks with at-most-bipartite nonlocal resources, even in the presence of globally shared classical randomness. A final step of the argument invokes convexity to show that if this linear constraint holds conditioned on a particular value of David's resource \textit{outputs}, then it holds conditioned on his overall measurement \textit{outcome} $D$ which is a function of the resource outputs.

The above result completes the demonstration that entangled measurements on bipartite states can provide a quantum advantage over local wirings of bipartite nonsignaling resources, even if the nonsignaling resources are potentially super-quantum like the PR box and globally shared classical randomness is allowed. The existence of such an advantage was previously unknown in networks where all pairs of parties are allowed to share the bipartite resources. Intriguingly, this example suggests a new way of expressing/realizing the power of a \textit{three}-way entangled measurement---specifically, the GHZ-basis measurement---for outperforming wirings of PR boxes. In contrast, paradigmatic correlations involving the \textit{bipartite} Bell basis measurement, which otherwise device-independently certify an entangled measurement under the stronger assumption that the experiment admits a valid quantum description, can be simulated by various arrangements of wired PR boxes as described in \cite{bierhorst:2023}.

Notably, the above result can be generalized to rule out higher-order resources in the $\textsf{NSW}$ paradigm as follows: 

\begin{proposition}\label{prop:QEM_notNSW}
For any fixed choice of $K\ge2$, there is a $(K+2)$-party behavior that \emph{can} be achieved in the \textsf{QEM} paradigm with shared at-most-bipartite entangled quantum states and allowing for entangled measurements, that \emph{cannot} be achieved in a $(K+2)$-partite \textsf{NSW} network sharing at-most-$K$ partite nonsignaling resources where measurements are restricted to local wirings, even allowing for super-quantum resources and globally shared classical randomness.
\end{proposition}

\noindent The explicit scenario described previously corresponds to the $K=2$ case of this proposition. For the general case, the \textsf{QEM} behavior is obtained in a star network where a central party shares one Bell pair with each of the other $K+1$ parties (so a total of $K+1$ Bell pairs); the central party, corresponding to ``David" in the $K=2$ special case, then performs a projection measurement onto the $(K+1)$-qubit GHZ state $(\ket{00\cdots0}+ \ket{11\cdots 1})/\sqrt 2$. Conditioned on the success of this measurement, which occurs with probability $1/2^{K+1}$, the other $K+1$ parties possess the higher-order GHZ state; this can then be used to perform a LOSR-GMNL test that witnesses genuine $(K+1)$-partite nonlocality, such as that of equation (5) in \cite{mao:2022} or similar constraints in \cite{cao:2022} and \cite{coiteux:2021a}. These tests, which are noise-robust, rule out underlying networks of at-most-$K$-partite nonsignaling resources. The achievable correlations for \textsf{QEM} in the higher $K>2$ cases are notable for not actually requiring $K$-partite quantum resources, just bipartite ones, while ruling out up-to-$K$-partite wired nonsignaling boxes.

The generalized impossibility proof for \textsf{NSW} follows directly from a careful read-through of the Appendix \ref{s:app} argument for $K=2$, making a note to replace $A,B,C$ with $K+1$ Alices $A_1,...,A_{K+1}$ with respective settings $X_1,...,X_{K+1}$, and David is the $(K+2)$th player; meanwhile the allowed resources become $K$-partite instead of bipartite. The only notable difference is that conditioned on David's success outcome, the resources he previously shared with other parties collapse to $(K-1)$-partite resources, which for $K=2$ are local random variables but for $K>2$ can be nonclassical resources shared among the other parties. However, these residual $(K-1)$-partite resources, in combination with any already-present $K$-partite resources shared by the non-David parties, cannot result in a distribution that is $(K+1)$-partite LOSR-GMNL. 

\section{Interpreting the result in paradigms  disallowing globally shared classical randomness, or in a restricted network topology}

Our results demonstrate that for $K\ge 2$, the \textsf{QEM} paradigm can outperform \textsf{NSW} in networks of $K+2$ parties where nonclassical resources are allowed for any subset of at most $K$ parties and globally shared classical randomness is also permitted. However, it is noteworthy that the witnessing \textsf{QEM} behavior only uses bipartite resources arranged in a star network with one central party individually connected to the other $K+1$ parties, in particular not requiring more-than-bipartite resources even for values of $K>2$, nor does it use the allowed globally shared classical randomness. The \textsf{QEM}/\textsf{NSW} separation can thus be witnessed in more restrictive scenarios. Examining these provides some broader context for highlighting the power of the \textsf{QEM} behavior.

\medskip

\noindent\textit{No globally shared classical randomness}. Removing the possibility of globally shared classical randomness enables the most natural generalization of our results downward to the $K=1$ and $K=0$ cases. These generalizations have some mathematical appeal as canonical base cases of a ladder of results, although their practical relevance appears limited. To elaborate, let us enlist the behavior of Figure \ref{f:scheme} to witness a \textsf{QEM}/\textsf{NSW} separation under the following minor reformulation: 

\medskip

\noindent``Networks of $K+2$ parties sharing bipartite quantum resources and allowing entangled measurements can exhibit behaviors unattainable by networks of $K+2$ parties sharing at-most-$K$-partite nonclassical resources that are restricted to wirings." 

\medskip

\noindent The key modifications are that the \textsf{QEM} paradigm now uses specifically \textit{bipartite} resources for all values of $K$, and globally shared classical randomness is not permitted in either the \textsf{QEM} paradigm or the \textsf{NSW} case. Our results in the previous section still witness the \textsf{QEM}/\textsf{NSW} separation under the above reformulation for $K\ge 2$, because the \textsf{QEM} behavior of Figure \ref{f:scheme} still meets the stated criteria while the impossibility proof for \textsf{NSW} in the previous section applies immediately to a more-restrictive paradigm that disallows globally shared classical randomness. Moreover, for $K=1$ we can now take the natural downward extension of our \textsf{QEM} behavior to be as follows: a Bell basis measurement is performed by party $B$ of the $(K+2)$-partite configuration in Figure \ref{f:tripart}(a) comprising two Bell pairs shared with $A$ and $C$, respectively; conditioned on ``success'', which now is projection onto the two-party ``GHZ'' state $(\ket{00}+\ket{11})/\sqrt 2$, $A$ and $C$ can perform a standard 2-party Bell test (such as the CHSH test) to rule out a network of 1-partite nonsignaling resources---recalling $K$ equals one---which are nothing more than unshared classical random variables. Here, the crucial difference from the earlier sections is that \textsf{QEM} behaviors are still allowed \textit{bipartite} resources; even if \textsf{NSW} scenarios were allowed globally shared classical randomness in the $K=1$ case, it would still not enable the violation of Bell inequalities. Incidentally, an entanglement swap isn't necessary to exhibit this somewhat trivial separation, as the power of bipartite quantum resources over classical random variables is witnessed by any Bell test; it is just that the swap configuration is the most direct parallel to the $K\ge 2$ \textsf{QEM} examples studied earlier. However it bears mentioning that depending on the experimental architecture, entanglement swapping \textit{can} be the most straightforward approach for witnessing nonclassicality, such as in the heralded loophole-free Bell experiment of \cite{hensen:2015} employing nitrogen vacancy centers.

Finally, moving down to the $K=0$ case, the \textsf{QEM} scenario reduces to two parties sharing a single Bell state, where the ``swapping" party performs the  one-party ``GHZ" measurement which is a single-qubit measurement in the $\{\ket +,\ket -\}$ basis; the other party can rule out correlations enabled by $0$-partite resources---which must be \textit{constant} for any choice of measurement setting in the absence of even 1-partite classical random variables---by simply measuring in any basis other than the $\{\ket +,\ket -\}$ basis to generate probabilistic output. 

The downward generalizations described above are primarily of academic interest, since it is well understood that \textit{bipartite} quantum resources can of course provide an advantage over 1- and 0-partite nonsignaling resources. An arguably more interesting question, returning to the $K\ge 2$ cases, is whether other \textsf{QEM} behaviors, perhaps simpler ones, can witness the separation in a \textsf{NSW} paradigm that now prohibits globally shared classical randomness. Such a separation has been previously considered in Reference \cite{renou:2019}, which introduced a tripartite behavior argued to exhibit a novel form of genuine quantum nonlocality. That behavior, which utilizes entangled measurements on bipartite-entangled Bell states arranged in the triangle network as in Figure \ref{f:tripart} (b), does not involve measurement settings and so is classically replicable in the presence of globally shared randomness. However, in the absence of this resource, the authors of \cite{renou:2019} expressed a belief that the behavior cannot be obtained with PR boxes, though this remains unproven.

To further examine the matter, let us examine a \textsf{QEM} behavior that is inspired by, but simpler than, the behavior of Figure \ref{f:scheme}. This behavior might be considered a natural candidate for something impossible within \textsf{NSW} \textit{without} globally shared randomness, but it turns out to in fact be possible within the paradigm. The result illustrates the importance of careful argumentation regarding \textsf{NSW}, since the basic intuition reflected in Figures \ref{f:bundle} and \ref{f:choppedbundles} here leads to a flawed conclusion if taken too far, as alluded to in the previous section.

The example we consider, for the $K=2$ case, is whether the parties Alice/Bob/Charlie can witness a shared classical correlation of the form \begin{equation}\label{e:cointoss}
P(A=B=C=+1)=P(A=B=C=-1)=1/2,
\end{equation}
conditioned on a success outcome from David. The above correlation, while simple, is known to \textit{not} be possible in a tripartite network in which bipartite-only nonsignaling resources are allowed (with the prohibition of tripartite resources extending to a prohibition on globally shared classical randomness, which would trivially enable the correlation), as is demonstrated in Examples 1 and 4 of Ref.~\cite{wolfe:2019}. Now in the \textsf{QEM} paradigm, David can swap a GHZ state to $A$/$B$/$C$, enabling the correlation of \eqref{e:cointoss} \textit{conditioned on David's success outcome} if Alice, Bob, and Charlie each measure the resulting GHZ state in the computational basis. Notably, this $A/B/C$ approach is simpler than the previously studied LOSR-GMNL test required to violate \eqref{e:mao}, and the correlation of \eqref{e:cointoss} could have potential utility through, e.g., enabling shared conference cryptographic key that generalizes the one-time pad to a multi-party configuration \cite{epping:2016}. Turning to the \textsf{NSW} paradigm, intuition may suggest that achieving this behavior is impossible, by over-interpreting the illustration of Figure \ref{f:choppedbundles} which suggests $A$/$B$/$C$ possess only 1-partite and 2-partite resources conditioned on David's outcome. This is however misleading; Figure \ref{f:choppedbundles} reflects the state of affairs given a specific fixed choice of observed single output for each one of David's resources, but if David bins multiple such resource-output possibilities together to form a composite ``success" event, the resulting mixed behavior can exhibit the correlation of \eqref{e:cointoss} even if the correlations conditioned on the atomic events do not. The key difference from the earlier impossibility argument is that \eqref{e:mao} is linear in probabilities---and so its satisfaction is maintained over convex mixtures---whereas the condition \eqref{e:cointoss}, as well as its generalization to the inequality in Example 4 of \cite{wolfe:2019}, are not. 

For an explicit example of how the correlation is possible in the \textsf{NSW} paradigm, David can share a bipartite classical random variable with Alice with the distribution $P(A=D=+1) = P(A=D=-1)=1/2$, while also sharing two independent identical resources with Bob and Charlie; conditioned on \textit{either} observing $D=+1$ for all three resources, or $D=-1$ for all three resources, David reports ``success", which will occur with total probability $1/4$---notably, this actually exceeds the success probability of the GHZ swap in the \textsf{QEM} strategy which is only 1/8. Then, Alice/Bob/Charlie report the value of their resources; they will observe the correlation of \eqref{e:cointoss}.

\medskip

\noindent\textit{Consideration of the star network}. We conclude our examination of alternate paradigms by restricting to the star network in which a central party shares bipartite resources with all other parties, but no two non-central parties share bipartite resources. For $K=1$, this is the configuration of Figure \ref{f:tripart} (a), which contrasts with the fully connected networks as in Figure \ref{f:tripart} (b) for $K=1$ or Figure \ref{f:fourparty} for $K=2$. When considering the star network, the restriction to bipartite-only resources, as opposed to $K$-partite, applies to both \textsf{QEM} and \textsf{NSW} paradigms for all higher values of $K$. Globally shared classical randomness can be allowed; it does not provide any utility for \textsf{NSW} schemes in the discussion below. 

The GHZ swap behavior of Figure \ref{f:scheme} is still possible in the \textsf{QEM} paradigm, as it only ever used the resources present in a star network. What is now different is that one can prove impossibility of this behavior in the now-more-restricted \textsf{NSW} paradigm by invoking earlier results on the impossibility of nonlocality swapping. In particular, \textsf{QEM}/\textsf{NSW} separation in the $K=1$ case---where the \textsf{QEM} behavior comprises a Bell basis measurement by the central party, which enables conditioned-on-success Bell violations by the outer parties---is directly implied by the results in references \cite{short:2006,chao:2017,Short_2010} as discussed in the introduction, and so does not require a more general argument in  the style of Appendix \ref{s:app}. 

Generalization of the \textsf{QEM}/\textsf{NSW} separation argument to $K\ge 2$ cases can be approached in two different ways. First, one can degenerately use the $K=1$ argument to describe behaviors possible for \textsf{QEM} and impossible for \textsf{NSW}, using only a subset of the $K+2$ parties, since the three-party star network of Figure \ref{f:tripart} (a) is a subnetwork of the higher-order star networks---and since the entanglement swap behavior is impossible in the subnetwork, it will be impossible as a sub-behavior of the full network. Thus the central party (David for $K=2$) does not need to perform the full GHZ basis measurement to rule out the \textsf{NSW} paradigm in star networks of $K+2$ parties for $K\ge 2$, and can instead achieve this purpose by just performing a basic Bell-basis entanglement swap on a two-element subset of his qubits while the two corresponding outlying parties perform a Bell-CHSH test. 

A more interesting second approach to generalization recognizes that the conditioned-on-success LOSR-GMNL behaviors 
are more exotic/powerful effects than ``merely" bipartite nonlocality, and thus they more fully exhibit the power of the \textsf{QEM} paradigm over \textsf{NSW}. That is, there is an additional quantum advantage available to quantum-entangled measurements, going beyond the basic circumvention of the impossibility of nonlocality swapping constraining \textsf{NSW}. \textsf{QEM}'s ability to swap LOSR-GMNL correlations can be considered a \textit{strictly} stronger effect than standard entanglement swapping in the following sense: the ability to violate \eqref{e:mao} in a three-party experiment implies the ability to violate a CHSH inequality in a two-party experiment, because if parties $A$ and $C$ are co-located and collapsed into one party, the first four terms of \eqref{e:mao} can be converted into a standard CHSH expression if $C$ is supplied the setting $Z=1$ and $A/C$ apply an appropriate map of their joint outcomes to a single outcome. This CHSH expression necessarily violates the classical bound of 2 if the bound of 4 in \eqref{e:mao} is violable because the fifth term of \eqref{e:mao} has an algebraic maximum of 2. Consideration of this example in the star network thus highlights a stronger entangled-measurement effect enabled by the \textsf{QEM} paradigm, going further beyond the limitations upon \textsf{NSW} imposed by the impossibility of nonlocality swapping alone, and so generalizes the results of \cite{short:2006,chao:2017,Short_2010} to higher order nonlocal effects.


\section{Conclusion}

We have shown that in fully connected networks, entangled measurements on bipartite quantum resources enable behaviors that are inaccessible with non-entangled measurements on bipartite and higher-partite nonsignaling resources, even if the nonsignaling resources are super-quantum and globally shared classical randomness is allowed. The quantum-achievable behaviors are robust and thereby experimentally testable. Such an experimental test can be considered a device-independent demonstration, in the sense that the single assumption of no-more-than-bipartite resources allows the inference of the presence of quantum-entangled measurements directly from the experimental statistics, without detailed assumptions of the inner workings of experimental devices. This inference is immediately valid in ruling out the special case of unentangled measurements on specifically \textit{quantum} resources such as in the paradigms of \cite{bierhorst:2023} and \cite{supic:2022}, allowing a practically useful inference within these more physically motivated scenarios.


Our method exhibits the multipartite character of the GHZ-basis entangled measurement by enabling a witness of genuine multipartite nonlocality according to the LOSR-GMNL paradigm. Future work relating this feature to a conceptualization of ``genuinely multipartite entangled measurements'', paralleling the new definitions of LOSR-GMNL \cite{coiteux:2021} or closely related LOSR notions of genuinely multipartite entanglement \cite{navascues:2020,schmid:2023}, will enhance our understanding of entangled measurements. 
More concretely, our results leave open an interesting technical matter as to whether \textit{three} parties (instead of four) can witness the advantage of quantum-entangled measurements of bipartite quantum resources over local wiring measurements of general nonsignaling bipartite resources in a fully connected network, or more generally whether $K+1$ parties (instead of $K+2$) performing entangled measurements on $K$-partite quantum resources can observe behaviors inaccessible to local wirings of $K$-partite nonsignaling resources. To address this, one could explore adding an additional measurement setting for the swapping party to elevate the behavior of Figure \ref{f:scheme} to a 4-party, 2-settings-for-each-party, 2-outcome behavior, a scenario admitting the powerful simplification technique of Jordan's lemma. Such behaviors could potentially generalize results of \cite{bierhorst:2023} by using David's extra measurement setting to pursue, for example, alignment of measurement outcomes with one of the other parties, which could interplay with the basic Figure \ref{f:scheme} behavior to witness an advantage of entangled measurements over unentangled measurements on \textit{tripartite} quantum resources in a four-party network, or even lead to the stronger result of witnessing the \textsf{QEM}/\textsf{NSW} separation for four parties sharing tripartite nonclassical resources. Continuing work in this area will refine our fundamental understanding of entangled measurements, which is of timely relevance given the central role entangled measurements will play in realizing the full capabilities of near-future emerging quantum networks.

\medskip

\noindent\textbf{Acknowledgments.} The authors thank Elie Wolfe and Rob Spekkens for the insightful suggestion of the quantum-achievable behavior described in Figure \ref{f:scheme} as a candidate for something that would not be possible for local wirings of  nonsignaling resources, along with the suggestion of the generalization to $K\ge2$ cases. The authors also thank Sujit Mainali and Nicholas Fore for help designing certain figures. This work was partially supported by NSF grants \#2210399 and \#232880.

\appendix
\section{Spoofing Entanglement Swapping with Bipartite Resources}\label{s:spoofing}

It is not hard to spoof the entanglement swapping protocol without actually employing an entangled measurement, if the swapped-to parties are allowed to share pre-existing bipartite quantum resources, and globally shared classical randomness is permitted. In particular, if we make Bob the swapping party following Figure \ref{f:tripart} (a), then as discussed in the comments preceding Equation (16) of \cite{bancal:2018}, the three parties can spoof the protocol by sharing the (unnormalized) classical-quantum state \begin{equation}\label{e:cq}
\sum_{k=0}^3\ket{k}\!\bra{k}_B\otimes\ket{\phi_k}\!\bra{\phi_k}_{AC}
\end{equation}
where Bob possesses the classical register $\ket k\!\bra{k}$ and the states $\ket{\phi_k}\!\bra{\phi_k}$ are the four Bell states indexed by $k=\{0,1,2,3\}$. By measuring his state, Bob learns which resulting Bell state is possessed by Alice and Charlie; this corresponds to what they would have possessed under the standard entanglement swapping protocol, but now Bob does not have to make an entangled measurement. To make clear the usage of only bipartite quantum resources and classical randomness, we can interpret \eqref{e:cq} as resulting from Alice and Charlie possessing one copy of each of the four Bell states, and then all three parties consult a shared four-output classical random variable; Bob reports the result as his measurement outcome $k$, while Alice and Charlie use their knowledge of $k$ to select their $k$th Bell state for any measurement they may subsequently perform.

Interestingly, globally shared classical randomness is not required to spoof the entanglement swapping procedure. Using the numbering 
\begin{equation*}
\ket{\phi_0}=\frac{\ket{00}+\ket{11}}{\sqrt 2} \qquad
\ket{\phi_1}=\frac{\ket{00}-\ket{11}}{\sqrt 2} \qquad
\ket{\phi_2}=\frac{\ket{01}+\ket{10}}{\sqrt 2} \qquad
\ket{\phi_3}=\frac{\ket{01}-\ket{10}}{\sqrt 2} 
\end{equation*}
let Alice and Charlie initially share the Bell state $\ket{\phi_0}=\ket{\Phi^+}$, and let (only) Alice and Bob share access to a uniformly distributed four-output classical random variable. Note Alice and Bob could implement this quantumly, by sharing two parallel Bell states and measuring them in the computational basis. Then based on the output $k$ of the four-output random variable, Alice applies to her share of the state $\ket{\phi_0}_{AC}$ the local unitary $I$ (if $k=0$), $Z$ ($k=1$), $X$ ($k=2$), or  $ZX$ ($k=3$). These are the local unitary corrections used in quantum teleportation, where $I$ is the identity and $Z$ and $X$ are Pauli matrices. With Bob reporting $k$ as his measurement outcome, this will result in Alice and Charlie possessing the correct post-swap Bell state, and Charlie did not need to know the output of the classical random variable shared only by Alice and Bob.

The above shows how Alice and Charlie can replicate the \textit{quantum states} resulting from the entanglement swapping protocol. Alternatively, we can view entanglement swapping from the perspective of the \textit{observed behavior}; here, the standard device-independent implementation to witness entanglement swapping is as follows (see e.g.~\cite{chao:2017,renou:2018}): Alice and Charlie measure their states with the respective Bell-CHSH measurements $A_0=Z$, $A_1=X$, and $C_{0,1}=\frac{1}{\sqrt 2}(X\pm Z)$; then in the honest implementation, Alice and Charlie's measurements result in a maximal violation of a symmetry of the CHSH inequality that depends on the result of Bob's entanglement swap. These symmetries are listed preceding Theorem 1 of Ref.~\cite{renou:2018}, which we reproduce here:
\begin{align*}
&\textnormal{CHSH}_0=\langle A_0C_0\rangle+\langle A_0C_1\rangle+\langle A_1C_0\rangle-\langle A_1C_1\rangle\\
&\textnormal{CHSH}_1=\langle A_0C_0\rangle+\langle A_0C_1\rangle-\langle A_1C_0\rangle+\langle A_1C_1\rangle\\
&\textnormal{CHSH}_2=-\langle A_0C_0\rangle-\langle A_0C_1\rangle+\langle A_1C_0\rangle-\langle A_1C_1\rangle\\
&\textnormal{CHSH}_3=-\langle A_0C_0\rangle-\langle A_0C_1\rangle-\langle A_1C_0\rangle+\langle A_1C_1\rangle
\end{align*}
Note that maximization of these quantities, with the choice of quantity to be maximized depending on Bob's outcome, corresponds to the three-party win condition represented by equation (1) in \cite{chao:2017} (there the roles of Bob and Charlie are reversed). Alice can use the strategy of the preceding paragraph to maximize each of the above CHSH quantities; however, this requires Alice to perform quantum manipulations on her state prior to measuring it, whereas a more basic strategy is as follows. Alice and Charlie still share $\ket{\phi_0}=\ket{\Phi^+}$ and now always measure it according to the measurements $A_0=Z$, $A_1=X$, and $C_{0,1}=\frac{1}{\sqrt 2}(X\pm Z)$; Alice once again shares a four-output random variable with (only) Bob and depending on its output, she selectively flips some of her CHSH measurement outcomes. Specifically, if $k=0$ she reports her outcome directly, if $k=1$ she flips her outcome when she is observing $A_1$, for $k=2$ she flips $A_0$, and for $k=3$ she flips both. This gives the corrections necessary to maximize all $\textnormal{CHSH}_k$ quantities listed above. Importantly, Alice is now effectively using her quantum state shared with Charlie as a black box whose outputs she classically post-processes depending on the output of the resource shared with Bob. In particular, this is an explicit \textsf{NSW} protocol that uses bipartite-only nonsignaling resources (without even resorting to super-quantum PR boxes) and does not require globally shared classical randomness.

\section{Noise robustness and GHZ-basis resolution of the central node measurement}\label{s:resolution}
Consider a four-party star network as in the first part of Figure~\ref{f:scheme}, where the three branch-parties Alice, Bob, and Charlie independently share a noisy bipartite quantum state with the central party David. The central party performs an entangling measurement (in the GHZ basis) on each share of the bipartite state that it independently shares with Alice, Bob, and Charlie. Then tracing out the central node system results in a post-measurement state shared by the three branch-parties on which they can perform a test of genuine nonlocality in networks---for instance, the LOSR-GMNL test. We want to study the noise robustness of such a protocol from the quantum violation of the three-party version of the Mao inequality \eqref{e:mao} which is known to serve as a linear witness for LOSR-GMNL~\cite{mao:2022}. There are different ways to model the noise associated with the quantum resources; here we adopt a simple noise model with all three pre-shared bipartite states as two-qubit Werner states, i.e., $\rho_{AD_1}$ (for Alice-David), $\rho_{BD_2}$ (for Bob-David), and $\rho_{CD_3}$ (for Charlie-David) as shown below: 
\begin{equation}\label{eq:rep2}
\rho_{XD_i} = F\Phi^+ + \frac{1-F}{3}(I-\Phi^+), 
\end{equation}
where $\Phi^+\equiv|\Phi^+\rangle\langle\Phi^+|$, $|\Phi^+\rangle=(|00\rangle+|11\rangle)/\sqrt{2}$, and $(X,i)\in\{(A,1),(B,2),(C,3)\}$. The parameter $F$ in~\eqref{eq:rep2} is a non-negative scalar quantity which denotes the fidelity value with respect to the state $|\Phi^+\rangle$~\cite{Jozsa01121994}, so we have $F = \langle\Phi^+|\rho_{XD_i}|\Phi^+\rangle$.

We now proceed with the robustness analysis. The initial state of the four-party system, before the central party performs a measurement, can be expressed as $\rho_{ABC\bar{D}}=\rho_{AD_1}\otimes\rho_{BD_2}\otimes\rho_{CD_3}$, where $\bar{D}\equiv D_1D_2D_3$. The central party performs a GHZ-basis measurement, the complete set of which is shown below: 
\begin{eqnarray*}
|\mathrm{GHZ}_{0,1}\rangle = \frac{1}{\sqrt{2}}(|000\rangle\pm |111\rangle), \quad |\mathrm{GHZ}_{2,3}\rangle = \frac{1}{\sqrt{2}} (|010\rangle \pm |101\rangle),\\
|\mathrm{GHZ}_{4,5}\rangle = \frac{1}{\sqrt{2}}(|011\rangle\pm |100\rangle), \quad 
|\mathrm{GHZ}_{6,7}\rangle = \frac{1}{\sqrt{2}}(|110\rangle \pm |001\rangle).
\end{eqnarray*}
Table~\ref{tab:GHZ_like_states} below lists the local unitaries that can be applied to the canonical GHZ state $|\mathrm{GHZ}_0\rangle$ to obtain the remaining states of the eight-element set of GHZ basis states.
\begin{table}[!htb]
\caption{\label{tab:GHZ_like_states}GHZ-like states obtained with local unitaries on the canonical GHZ state $|\mathrm{GHZ}_{0}\rangle$.}
\centering
\begin{tabular}{c}
\hline\hline
$(I_A\otimes I_B\otimes Z_C)|\mathrm{GHZ}_{0}\rangle = |\mathrm{GHZ}_{1}\rangle$\\
$(I_A\otimes X_B\otimes I_C)|\mathrm{GHZ}_{0}\rangle = |\mathrm{GHZ}_{2}\rangle$\\
$(I_A\otimes X_B\otimes Z_C)|\mathrm{GHZ}_{0}\rangle = |\mathrm{GHZ}_{3}\rangle$\\
$(I_A\otimes X_B\otimes X_C)|\mathrm{GHZ}_{0}\rangle = |\mathrm{GHZ}_{4}\rangle$\\
$(Z_A\otimes X_B\otimes X_C)|\mathrm{GHZ}_{0}\rangle = |\mathrm{GHZ}_{5}\rangle$\\
$(X_A\otimes X_B\otimes I_C)|\mathrm{GHZ}_{0}\rangle = |\mathrm{GHZ}_{6}\rangle$\\
$(X_A\otimes X_B\otimes Z_C)|\mathrm{GHZ}_{0}\rangle = |\mathrm{GHZ}_{7}\rangle$\\
\hline\hline
\end{tabular}
\end{table}
Suppose now that David performs the projective measurement $\Pi_0 = \ket{\mathrm{GHZ}_0} \bra{\mathrm{GHZ}_0}$, then the normalized post-measurement state (upon success), $\rho_{ABC}^{(0)}=\mathrm{Tr}_{\bar{D}}\left[(I_{ABC}\otimes\Pi_0)\rho_{ABC\bar{D}}\right]$, shared between Alice, Bob and Charlie, is expressible as
\begin{equation}\label{eq:Post_Meas_State1}
\rho_{ABC}^{(0)} = \frac{1}{8p_0}\sum_{\mu,\nu,\lambda=0}^3 q_1(\mu)q_2(\nu)q_3(\lambda)(\sigma_{\mu}\otimes\sigma_{\nu}\otimes\sigma_{\lambda})\Pi_{0}(\sigma_{\mu}\otimes\sigma_{\nu}\otimes\sigma_{\lambda}),
\end{equation}
where $q_i(0)=F$, $q_i(1)=q_i(2)=q_i(3)=(1-F)/3$, and $p_0 = \mathrm{Tr}[\rho_{ABC}^{(0)}]$ denotes the probability of obtaining the outcome corresponding to $|\mathrm{GHZ}_0\rangle$ which is $\frac{1}{8}$. The matrices $\sigma_0,\sigma_1,\sigma_2,\sigma_3$ are the Pauli matrices $I,X,Y,Z$, respectively. One can also evaluate that an outcome corresponding to any other GHZ-measurement $\Pi_i=|\mathrm{GHZ}_i\rangle\langle\mathrm{GHZ}_i|$ for $1\le i\le 7$ arises with the same probability $p_i = \frac{1}{8}$, and the post-measurement state is the same as~\eqref{eq:Post_Meas_State1} with $\Pi_0$ replaced with $\Pi_i$ for $i\in\{1,\ldots,7\}$. Because applying any local Pauli operator $\sigma_\mu\otimes\sigma_\nu\otimes\sigma_\lambda$ to $\Pi_0$ in~\eqref{eq:Post_Meas_State1} maps it to another GHZ-projector $\Pi_i$ (up to a global phase), every term in the mixture is a projector onto one of the GHZ basis states. The coefficients $q_1(\mu)q_2(\nu)q_3(\lambda)$ are non-negative and sum to $1$, so the output is a convex combination of the GHZ basis projectors, i.e., \eqref{eq:Post_Meas_State1} can be re-expressed as:
\begin{equation}\label{eq:Post_Meas_State2}
\rho_{ABC}^{(0)} = \sum_{i=0}^7 t_i\Pi_i,\,\,\mathrm{where}\,\,\Pi_i=|\mathrm{GHZ_i}\rangle\langle\mathrm{GHZ}_i|,\,t_i\ge 0,\,\sum_{i=0}^7 t_i = 1.
\end{equation}
In~\eqref{eq:Post_Meas_State2}, the coefficients are $t_0=\frac{1}{2}(a^3 + b^3 + c^3)$, $t_1=\frac{1}{2}(a^3 + b^3 - c^3)$, and $t_2=\cdots=t_7=\frac{1}{2}ab$, where $a=(2F+1)/3$, $b=2(1-F)/3$, and $c=(4F-1)/3$.

Our task is to find the threshold for the fidelity $F$ such that the correlation obtained from measuring the state $\rho_{ABC|\Pi_0}$ using suitable measurements violates the Mao-Bell inequality, whose corresponding operator is:
\begin{equation}\label{eq:CanonicalMaoBellOp}
B_M = A_0\otimes(B_0 + B_1)\otimes I_C + A_1\otimes(B_0 - B_1)\otimes C_1 + 2A_0\otimes I_B\otimes C_0,
\end{equation}
where $A_{0,1}, B_{0,1}, C_{0,1}$ are Hermitian operators acting on the Hilbert spaces of Alice, Bob, and Charlie, respectively. The maximum expected value of~\eqref{eq:CanonicalMaoBellOp} for all correlations permissible in any setup where Alice, Bob, and Charlie perform local operations on their respective shares of bipartite resources (in addition to having access to globally shared classical randomness) is $4$. The maximum quantum value of $2(\sqrt{2}+1)$ for the expected value of~\eqref{eq:CanonicalMaoBellOp} is obtained from the measurements $A_0=C_0=Z$, $A_1=C_1=X$, and $B_{0,1}=\frac{1}{\sqrt{2}}(Z\pm X)$ on $|\mathrm{GHZ}_0\rangle$. When we use these measurements on the state in~\eqref{eq:Post_Meas_State2}, we get the expected value of:
\begin{equation}
\langle B_M\rangle_{\rho} = \sum_{i=0}^7 t_i\langle\mathrm{GHZ}_i|B_M|\mathrm{GHZ}_i\rangle=\frac{2}{27}(4F-1)^2(2\sqrt{2}F+3+\sqrt{2}),
\end{equation}
which leads to a violation if $\frac{2}{27}(4F-1)^2(2\sqrt{2}F+3+\sqrt{2})>4$ holds, thereby giving us a numerical lower bound for $F$ which is approximately $0.941$. 

Now suppose that instead of the two-outcome projective measurement $\{\Pi_0,I-\Pi_0\}$, David performs the full eight-outcome projective measurement $\{\Pi_i\}_{i=0}^7$, where $\Pi_i=|\mathrm{GHZ}_i\rangle\langle\mathrm{GHZ}_i|$ as before. Then depending on David's outcome, the state shared by Alice, Bob, and Charlie is projected onto a GHZ-type state. Since Alice, Bob, and Charlie are not privy to David's measurement in the honest implementation of the protocol, they perform the measurements tailored to the canonical GHZ state $|\mathrm{GHZ}_0\rangle$. The correlations obtained then violate a different symmetry of the Mao-Bell inequality. Since every other GHZ-like state is obtained with local unitaries $U_i\equiv U_A\otimes U_B\otimes U_C$ as shown in Table~\ref{tab:GHZ_like_states}, we have $\Pi_i=U_i\Pi_0 U_i^\dagger$. It is straightforward to show that following David's measurement of $\Pi_i$, the state shared by Alice, Bob, and Charlie is related to $\rho_{ABC}^{(0)}$ through the same local unitary, i.e., $\rho_{ABC}^{(i)}=U_i\rho_{ABC}^{(0)}U_i^\dagger$. Then performing the same measurements (tailored to the canonical GHZ state) on $\rho_{ABC}^{(i)}$ results in correlations that give the same expected value for a relabeled Mao-Bell expression whose corresponding operator is related to the canonical Mao-Bell operator through the same local unitary $U_i$, i.e.,
\begin{equation}
\mathrm{Tr}[B_{M,i}\Pi_i]=\mathrm{Tr}[U_iB_{M}U_i^\dagger U_i\Pi_0 U_i^\dagger]=\mathrm{Tr}[U_iB_M\Pi_0 U_i^\dagger]=\mathrm{Tr}[B_M\Pi_0],
\end{equation}
and so $\langle B_{M,i}\rangle_{\rho^{(i)}}=\langle B_M\rangle_{\rho^{(0)}}$,
where $B_{M,i}=U_iB_MU_i^\dagger$ is the relabeled version of the canonical Mao-Bell operator $B_M$. It follows that the critical threshold on the fidelity $F$ remains the same as for the canonical case (corresponding to $i=0$) which is $0.941$ approximately. Table~\ref{tab:Relabelled_MaoOperators} lists the relabeled Mao-Bell operators and expressions and the corresponding local unitaries.


\begin{table}[!htb]
\caption{\label{tab:Relabelled_MaoOperators}Relabeled versions of the canonical Mao-Bell operator in~\eqref{eq:CanonicalMaoBellOp}, whose corresponding Mao-Bell expression is $A_0(B_0+B_1)+A_1(B_0-B_1)C_1+2A_0C_0$. For brevity, we have omitted the tensor product symbol `$\otimes$' in the local unitaries and Bell operators, so $U_A\otimes U_B\otimes U_C\equiv U_A U_B U_C$.}
\centering
\resizebox{\textwidth}{!}{%
\begin{tabular}{ccc}
\hline\hline
Local unitary & Relabeled Bell operator & Relabeled Bell expression \\
\hline
$I_AI_BZ_C$ & $\sqrt{2}Z_AZ_BI_C - \sqrt{2}X_AX_BX_C + 2Z_AI_BZ_C$ & $A_0(B_0 + B_1)-A_1(B_0 - B_1)C_1 + 2A_0C_0$\\
$I_AX_BI_C$ & $-\sqrt{2}Z_AZ_BI_C + \sqrt{2}X_AX_BX_C + 2Z_AI_BZ_C$ & $-A_0(B_0 + B_1) + A_1(B_0 - B_1)C_1 + 2A_0C_0$ \\
$I_AX_BZ_C$ & $-\sqrt{2}Z_AZ_BI_C - \sqrt{2}X_AX_BX_C + 2Z_AI_BZ_C$ & $-A_0(B_0 + B_1) - A_1(B_0 - B_1)C_1 + 2A_0C_0$ \\
$I_AX_BX_C$ & $-\sqrt{2}Z_AZ_BI_C + \sqrt{2}X_AX_BX_C - 2Z_AI_BZ_C$ & $-A_0(B_0 + B_1) + A_1(B_0 - B_1)C_1 - 2A_0C_0$\\
$Z_AX_BX_C$ & $-\sqrt{2}Z_AZ_BI_C - \sqrt{2}X_AX_BX_C - 2Z_AI_BZ_C$ & $-A_0(B_0 + B_1) - A_1(B_0 - B_1)C_1 - 2A_0C_0$\\
$X_AX_BI_C$ & $\sqrt{2}Z_AZ_BI_C + \sqrt{2}X_AX_BX_C - 2Z_AI_BZ_C$ & $A_0(B_0+B_1) + A_1(B_0-B_1)C_1 - 2A_0C_0$ \\
$X_AX_BZ_C$ & $\sqrt{2}Z_AZ_BI_C - \sqrt{2}X_AX_BX_C - 2Z_AI_BZ_C$ & $A_0(B_0 + B_1) - A_1(B_0 - B_1)C_1 - 2A_0C_0$ \\
\hline\hline
\end{tabular}%
}
\end{table}

\section{Formal proof of the impossibility of witnessing the behavior in the \textsf{NSW} paradigm}\label{s:app}
As outlined in Section \ref{s:imposs}, the \textsf{QEM}-achievable 4-partite behavior of Figure \ref{f:scheme} is not possible to replicate in a network of at-most bipartite nonclassical resources under the \textsf{NSW} paradigm, even in the presence of globally shared classical randomness; we formally prove this here. 

Without loss of generality, we can assume a fixed value for any globally shared classical random variable; the argument below will apply for this fixed value, and then by convexity the derived constraint will hold for the average over all values of shared randomness. Such a reduction is intuitive but can also be treated formally as is done in Equation (18) of Ref.~\cite{bierhorst:2024}; we provide a little more detail on this formalization in \ref{s:factor} after concluding the main argument. In the main argument we now present, this simplification allows us to effectively consider a network comprising only bipartite nonsignaling resources \textit{without} globally shared classical randomness.

\subsection{Main proof}\label{s:mainproof}

Our proof uses the framework of \cite{bierhorst:2024}, which may be helpful to consult while following through the argument. In the \textsf{NSW} paradigm, David has a decision tree as in Figure 3 of \cite{bierhorst:2024} that governs the order in which he measures resources and what inputs he provides to them. David's path through the decision tree depends on the outputs he sees from his nonsignaling resources as he observes them, and he has different terminal ``end" locations he can arrive at depending on how his resources produce outputs; in particular, each ``end" location corresponds to a single distinct vector of observed resource outputs. David will associate certain end locations with ``success", mapping these to his final reported outcome of ``success", and others he will deem ``failure". 
Let $D$ be the random variable denoting the ``end" location---or equivalently, $D$ is the vector of David's observed resource outputs---and let $\mathcal S$ denote the set of $D$ outcomes $d$ that David bins together and maps to his overall measurement output ``success". $\mathcal S$ is a composite event comprising a disjoint union of atomic events $d$.

Our goal is to show that
\begin{equation*}
{P(ABC|XYZ,\textnormal{David reports ``success"}) = P(ABC|XYZ, D\in \mathcal S)}
\end{equation*}
will obey any Bell-like linear constraint of the form $\sum_{abcxyz} \beta_{abcxyz}P(abc|xyz)\le B$ such as \eqref{e:mao} that witnesses LOSR-GMNL (i.e., a constraint whose violation requires the presence of a three-way nonclassical resource, even if three-way shared classical randomness is allowed).

We first highlight how $P(ABC|XYZ, D\in \mathcal S)$ is a convex mixture of $P(ABC|XYZ, D=d)$ distributions where the $d$ are the atomic events comprising $\mathcal S$, to allow us to focus directly on deriving linear constraints for $P(ABC|XYZ, D=d)$ distributions. Abbreviating $\{D\in \mathcal S\}$ with $\mathcal S$ and $\{D=d\}$ with $d$, we have
\begin{align}
P(ABC|XYZ, \mathcal S) &=P^{-1}(\mathcal S|XYZ)P(ABC, \mathcal S|XYZ)\nonumber\\
&=P^{-1}(\mathcal S)\sum_{d\in \mathcal S}P(ABC, d|XYZ)\nonumber\\
&=P^{-1}(\mathcal S)\sum_{d\in \mathcal S}P(ABC|d,XYZ)P(d|XYZ)\nonumber\\
&=P^{-1}(\mathcal S)\sum_{d\in \mathcal S}P(ABC|d,XYZ)P(d)\label{e:cvxmix}
\end{align}
where the no-signaling principle for the overall distribution (see \cite{bierhorst:2024} Section 3.1) is used to justify removing conditioning on $XYZ$ from probabilities involving David only. By the above, if it were known that $P(ABC|d,XYZ)$ obeys a Bell-like linear constraint of the form $\sum_{abcxyz} \beta_{abcxyz}P(abc|d,xyz)\le B$ for each value of $d$, we would be able to write
\begin{align*}
\sum_{abcxyz}\beta_{abcxyz}P(abc|xyz, \mathcal S) &= \sum_{abcxyz}\beta_{abcxyz}P^{-1}(\mathcal S)\sum_{d\in \mathcal S}P(abc|d,xyz)P(d) \\
&=\sum_{d\in \mathcal S}\frac{P(d)}{P(\mathcal S)}\sum_{abcxyz}\beta_{abcxyz}P(abc|d,xyz)\\
&\le \sum_{d\in \mathcal S}\frac{P(d)}{P(\mathcal S)}B\\
&= B,
\end{align*} 
as the $d$ are disjoint events comprising $\mathcal S$.

Thus it is sufficient to show that the desired bound $B$ applies to each distribution $P(ABC|d,XYZ)$ that is conditioned on a David atomic event $d$. To demonstrate this, we now enlist the representation of the distribution $P$ given by Equation (9) of Ref.~\cite{bierhorst:2024}. Applied to the current context, we see (for example) in Figure \ref{f:bundle} there are $m=13$ bipartite resources $\{R_k\}_{k=1}^{13}$. Each will have a distribution like $R_{11}(b^{(11)}c^{(11)}|y^{(11)}z^{(11)})$ ($R_{11}$ being one of the boxes shared by Bob and Charlie) which we will notate $R_k(\mathcal O^k|\mathcal I^k)$ with $\mathcal O^k, \mathcal I^k$ representing the outputs/inputs of the pair of parties involved, with the particular pair of parties depending on the choice of resource $k$. Note that while David does not have an overall \textit{setting}, his individual boxes can have \textit{inputs} which we denote $w^{(k)}$ corresponding to a box output $d^{(k)}$. Denote the subset of the $m$ resources involving David with $\mathcal D \subseteq \{1,...,m\}$; then, for each fixed choice of $abcxyz$, we have
\begin{align*}
P(abc|d,xyz)&=P(abc,d|xyz)P^{-1}(d|xyz)\\
&=P(abc,d|xyz)P^{-1}(d)\\
&=P^{-1}(d)\prod_{k=1}^m R_k(\mathcal O_k|\mathcal I_k) \qquad\qquad\qquad\qquad \textnormal{(following eq.~(9) of \cite{bierhorst:2024})}\\
&=P^{-1}(d)\prod_{k\notin \mathcal D}R_k(\mathcal O_k|\mathcal I_k)\prod_{k\in \mathcal D}R_k(p^{(k)}, d^{(k)}|s^{(k)}, w^{(k)})\\
&=P^{-1}(d)\prod_{k\notin \mathcal D}R_k(\mathcal O_k|\mathcal I_k)\prod_{k\in \mathcal D}R_k(p^{(k)}| d^{(k)},s^{(k)}, w^{(k)})R_k(d^{(k)}|s^{(k)}, w^{(k)})\\
&=P^{-1}(d)\prod_{k\in \mathcal D}R_k(d^{(k)}|w^{(k)})\prod_{k\notin \mathcal D}R_k(\mathcal O_k|\mathcal I_k)\prod_{k\in \mathcal D}R_k^{d^{(k)}, w^{(k)}}(p^{(k)}|s^{(k)})\\
\end{align*}
where in the fourth line, $p^{(k)}$ is one of $a^{(k)},b^{(k)}, c^{(k)}$ (based on whoever shares resource $k$ with David) and $s^{(k)}$ is the corresponding $x^{(k)}$, $y^{(k)}$ or $z^{(k)}$; note also $R_k(d^{(k)}|s^{(k)}, w^{(k)}) = R_k(d^{(k)}|w^{(k)})$ above by no-signaling (of the resources), and $R_k^{d^{(k)}, w^{(k)}}(p^{(k)}|s^{(k)})$ is just a notational shorthand equivalent to $R_k(p^{(k)}| d^{(k)},s^{(k)}, w^{(k)})$ that highlights how, when conditioned on David's input and output $w^{(k)}$ and $d^{(k)}$, $R_k$ acts as a one-party resource for the other party. Now, $\prod_{k\in \mathcal D}R_k(d^{(k)}|w^{(k)})=P(d)$; this is the statement (``obvious" by no-signaling, but also formally provable in the manner of \cite{bierhorst:2024}; see Section \ref{s:fj})  
that David's marginal distribution $P(d)$ can be obtained locally by examining the path through his decision tree leading to $d$,  and determining the probability of this path while considering only the David-only marginal distribution of his resources as they are consulted one by one. Hence we get a cancellation, and so
\begin{equation}\label{e:usesdavid}
P(abc|d,xyz)=\prod_{k\notin \mathcal D}R_k(\mathcal O_k|\mathcal I_k)\prod_{k\in \mathcal D}R_k^{d^{(k)}, w^{(k)}}(p^{(k)}|s^{(k)}).
\end{equation}
Any now-one-player resource $R_k^{d^{(k)}, w^{(k)}}(p^{(k)}|s^{(k)})$ is nothing more than a local random variable (indeed, a constant if $R_k$ is a PR box). Bipartite resources involving non-David players are unchanged. The distribution $P(abc|d,xyz)$ given by \eqref{e:usesdavid} is precisely what you get from a network of at-most-bipartite resources by constructing the three-party distribution according to equation (9) of \cite{bierhorst:2024} as you work your way through Alice, Bob, and Charlie's decision trees (as depicted in Figure 3 of \cite{bierhorst:2024}) to fill in the appropriate resource inputs. Since there are only bipartite nonclassical resources and no three-way nonclassical resources, the distribution cannot violate any inequality witnessing LOSR-GMNL such as \eqref{e:mao}, and so the success-conditional behavior $P(ABC|XYZ, D\in \mathcal S)$, which is a convex mixture of the $P(abc|d,xyz)$ distributions, cannot violate this (linear) inequality either.

\subsection{Formal justification of an intuitive formula for marginal distributions}\label{s:fj}

Ref.~\cite{bierhorst:2024} contains a derivation of the result that the overall joint distribution $P(\cdot|\cdot)$ induced by a network of nonsignaling resources is nonsignaling, as used in the above derivation of \eqref{e:cvxmix}. However, it is only \textit{implicit} in that derivation that the reduced marginal distribution of the subset of parties (and in particular the marginal distribution of the single party David, which is what is relevant to the comments preceding equation \eqref{e:usesdavid}) can be calculated as follows: 1) re-conceptualize each resource $R(\cdot|\cdot)$ as being shared only by the subset parties, where the reduced distribution $R(\textnormal{subset party outputs}|\textnormal{subset party inputs})$ is well-defined via the no-signaling property of the $R(\cdot|\cdot)$ resources, and then 2) working through the (original) decision trees of the parties in the subset and consulting these now-reduced-to-the-subset resources, obtaining the distribution via equation (9) of \cite{bierhorst:2024}.

We now demonstrate \textit{explicitly} how this can be seen in \cite{bierhorst:2024}. Referring to expression (17) of that work, that sequence of equations overall yields the equivalence
{\small 
\begin{equation}\label{e:17}
\sum_{\mathbf a_1} \mathbb P(\mathbf a_1, \mathbf a_2,\ldots, \mathbf a_n| \mathbf x_1, \mathbf x_2,\ldots, \mathbf x_n)=\prod_{k\notin \mathcal R_1}R_k(\cdots|\cdots)\prod_{k\in \mathcal R_1} R_k(a_{k_2}^{(k)},\ldots,a_{k_{n_k}}^{(k)}|x_{k_2}^{(k)},\ldots,x_{k_{n_k}}^{(k)}),
\end{equation}
}
where we adapt the notation a bit to the current scenario such that the bolded $\mathbf x_i/\mathbf a_i$ represent the overall setting/final outcome of party $i$, to contrast with un-bolded $x_i^{(k)}/a_i^{(k)}$ which refer to inputs/outputs for individual $R_k$ resources. Above, $\mathcal R_1$ denotes the set of resources shared by party 1 and $k_{1}, ..., k_{n_k} \subseteq \{1,...,m\}$ enumerates the parties sharing resource $R_k$. The key aspect of \eqref{e:17} is that on the right side, the conditional probability expressions for resources $R_k$ with $k\in \mathcal R_1$ are reduced, not containing party 1 inputs and outputs. This allows \cite{bierhorst:2024} to demonstrate no-signaling of $P$ by observing the expression \eqref{e:17} does not depend on party 1's setting $\mathbf x_1$ as no resource inputs $x^{(k)}_1$ or outputs $a^{(k)}_1$ (which could potentially depend on $\mathbf x_1$) appear, and so the distribution of players $2$ through $n$---note the marginalization on the left side of \eqref{e:17}---does not depend on the setting of the first player. 

For our current purposes, we furthermore observe that equation \eqref{e:17} demonstrates a \textit{formula} for computing the marginal distribution
$$
P(\mathbf a_2,\ldots, \mathbf a_n|\mathbf x_2,\ldots, \mathbf x_n)
$$
of players 2 through $n$: all resources in the set $\mathcal R_1$ are reduced to eliminate inputs/outputs $a_{k_1}/x_{k_1}$ (those involving player 1), and the expression on the right hand side of \eqref{e:17} is precisely what you would get from \cite{bierhorst:2024} eq.~(9) for an $(n-1)$-party network by going through the original decision trees for parties 2 through $n$ to populate the $R_k(\cdot|\cdot)$ expressions while treating any resource involving player 1 as a smaller, marginalized non-player-1-involving resource whose distribution is given by the formula
\begin{align*}
&R_k(a_{k_2}^{(k)},...,a_{k_{n_k}}^{(k)}|x_{k_2}^{(k)},...,x_{k_{n_k}}^{(k)})\\
=&\sum_{a^{(k)}_{k_1}}^{(k)}R_k(a_{k_1}^{(k)}, a_{k_2}^{(k)},...,a_{k_{n_k}}^{(k)}|x_{k_1}^{(k)},x_{k_2}^{(k)},...,x_{k_{n_k}}^{(k)}) \quad \text{for an arbitrarily chosen }x_{k_1}
\end{align*}
We can then obtain the $(n-2)$-party distribution $P(\mathbf a_3, .... , \mathbf a_n|\mathbf x_3, .... , \mathbf x_n)$ by iterating the procedure, repeating this until we have a subset of any size; in particular, the subset including just David, as discussed just preceding \eqref{e:usesdavid}. Naturally the procedure does not depend on eliminating player 1, then player 2,...; any players can be eliminated in any order.

We can justify the above iteration fully formally with an inductive argument demonstrating that the final reduced distributions (of both resources $R$ and overall distribution $P$) that are obtained in the iterative step-by-step manner are equivalent to the distributions that are obtained by marginalizing over all the eliminated parties at once (while simultaneously removing the marginalized parties' settings from the conditioner via the no-signaling principle). Suppose we have shown that the distribution $P(\mathbf a_i, .... , \mathbf a_n|\mathbf x_i, .... , \mathbf x_n)$ is equal to the network distribution given by $\prod_k R_k(\cdot|\cdot)$ where any inputs/outputs of parties $\{1,...,i-1\}$ have been removed from the $R(\cdot|\cdot)$ expressions, leaving reduced resources (well-defined by the no-signaling principle) each shared by a subset of those remaining players with indices $\{i,..., n\}$, and the reduced resources are populated according to the original decision trees of the $\{i,..., n\}$ players. Then this itself is a valid network-of-nonsignaling-resources distribution, to which we can apply the \eqref{e:17} expression, now summing over $\mathbf a_i$ to get
\begin{multline}\label{e:17alt}
\sum_{\mathbf a_i} \mathbb P(\mathbf a_{i}, \mathbf a_{i+1}, .... , \mathbf a_n| \mathbf x_i, \mathbf x_{i+1}, ... , \mathbf x_n)=\\
\prod_{k\notin \mathcal R_i}R_k(\cdots|\cdots)\prod_{k\in \mathcal R_i} R'_k(a_{k_j+1}^{(k)},...,a_{k_{n_k}}^{(k)}|x_{k_j+1}^{(k)},...,x_{k_{n_k}}^{(k)}),
\end{multline}
where for all the $k\in \mathcal R_i$, the outputs/inputs of player $i$ (previously enumerated as $a_{k_j}^{(k)}/x_{k_j}^{(k)}$) have been removed leaving the reduced distribution defined as (following (5) in \cite{bierhorst:2024})
\begin{align}\label{e:Rprime}
&R'_k(a_{k_j+1}^{(k)},\ldots,a_{k_{n_k}}^{(k)}|x_{k_j+1}^{(k)},\ldots,x_{k_{n_k}}^{(k)})=\nonumber\\
&\sum_{a_{k_j}^{(k)}}R_k(a_{k_j}^{(k)},a_{k_j+1}^{(k)},\ldots,a_{k_{n_k}}^{(k)}|x_{k_j}^{(k)},x_{k_j+1}^{(k)},\ldots,x_{k_{n_k}}^{(k)}) \quad \text{for some fixed choice }x_{k_j}^{(k)}
\end{align}
with the remaining players in the $\{i+1,...,n\}$ set following their decision trees to populate the specific choices of inputs/outputs in their $R'$ resources. Here, we have added the prime to $R'$ to emphasize that it is obtained not directly from the original full-party resource distribution, but from the intermediate reduced resource distribution in \eqref{e:Rprime} with only one additional party. But the left hand side in \eqref{e:17alt} is equivalent to 
$P(\mathbf a_{i+1}, .... , \mathbf a_n| \mathbf x_{i+1}, ... , \mathbf x_n)$, and the right hand side is equivalent to $\prod R(\cdot|\cdot)$ with inputs/outputs of parties $1,...,i-1, i$ removed (from the original, full $R(\cdot|\cdot)$ expressions) while the remaining players in $i+1,...,n$ populate the now-one-party-further-reduced $R$ resources according to their original decision trees. Both of these equivalencies are justified by the ``iterative" nature of taking marginal distributions of nonsignaling distributions; i.e., for the LHS of \eqref{e:17alt} we have 
\begin{align*}
&\sum_{\mathbf a_i} \mathbb P(\mathbf a_{i}, \mathbf a_{i+1},\ldots, \mathbf a_n| \mathbf x_i, \mathbf x_{i+1},\ldots, \mathbf x_n)\\
=&\sum_{\mathbf a_i}\sum_{\mathbf a_1,\ldots,\mathbf a_{i-1} }\mathbb P(\mathbf a_{1},\ldots, \mathbf a_{i-1},\mathbf a_{i}, \mathbf a_{i+1},\ldots, \mathbf a_n| \mathbf x_{1},\ldots, \mathbf x_{i-1},\mathbf x_i, \mathbf x_{i+1},\ldots, \mathbf x_n)\\
&\qquad\qquad\quad \text{(for an arbitrary choice of }\mathbf x_{1},\ldots,\mathbf x_{i-1})\\
=&\,\mathbb P(\mathbf a_{i+1},\ldots, \mathbf a_n| \mathbf x_{i+1},\ldots,\mathbf x_n)
\end{align*}
\noindent with the first equality following from the formal definition of $\mathbb P(\mathbf a_{i}, \mathbf a_{i+1}, .... , \mathbf a_n| \mathbf x_{i}, ... , \mathbf x_n)$ according to equation (5) of \cite{bierhorst:2024} and the second from the corresponding formal definition of $\mathbb P(\mathbf a_{i+1}, .... , \mathbf a_n| \mathbf x_i, \mathbf x_{i+1}, ... , \mathbf x_n)$. For the right side of \eqref{e:17alt} we apply the same reasoning to the $R'$ as defined in \eqref{e:Rprime}.

\subsection{Factoring out globally shared classical randomness}\label{s:factor}

If we collect any/all globally shared classical randomness into a single variable $\lambda$, then according to equation (18) of \cite{bierhorst:2024} and the surrounding discussion therein, the distribution $P(ABCD|XYZ)$ equals $\sum_\lambda P(\lambda) P_\lambda(ABCD|XYZ)$ where $P_\lambda(ABC|XYZ)$ is the distribution that results when the shared classical random variable takes the fixed value $\lambda$. $P_\lambda(ABC|XYZ)$ is equivalent to a distribution induced by (only) bipartite nonlocal resources (without shared classical randomness); since the argument in Section \ref{s:mainproof} that leads to \eqref{e:usesdavid} assumes a distribution of this form, that argument really shows that the Bell-type inequality applies to each $P_\lambda (ABC|XYZ,\mathcal S)$ for fixed choices of $\lambda$. But Bell functions are linear in the probabilities, and so if a linear constraint holds for each $P_\lambda$ it will apply to the convex mixture, as already observed in the arguments surrounding \eqref{e:cvxmix}. More explicitly, we can use the fact that $P(abc|xyz,\mathcal S) = P(abc,\mathcal S|xyz)P^{-1}(\mathcal S)$ for David's success event $\mathcal S$ (which doesn't depend on $XYZ$), along with the implications of equation (18) of \cite{bierhorst:2024} just mentioned, to manipulate a Bell-type expression as 
\begin{align*}
\sum_{abcxyz} \beta_{abcxyz}P(abc|xyz, \mathcal S) 
&= \sum_{abcxyz} \beta_{abcxyz}P^{-1}(\mathcal S)\sum_\lambda P(\lambda) P_\lambda (abc|\mathcal S, xyz)P_\lambda(\mathcal S)\\
&=P^{-1}(\mathcal S)\sum_\lambda P(\lambda) P_\lambda(\mathcal S)\sum_{abcxyz} \beta_{abcxyz}P_\lambda (abc|\mathcal S, xyz)\\
&\le P^{-1}(\mathcal S)\sum_\lambda P(\lambda) P_\lambda(\mathcal S)B= B
\end{align*}
so that satisfaction of the bound $B$ for each $P_\lambda$ implies satisfaction of the bound for the overall $P$.


\printbibliography

\end{document}